\documentclass[11pt,twoside]{article}

\usepackage{asp2004}
\usepackage{graphicx}

\begin{document}

\title{Effects of Rotation on Mass Loss for Population III stars}
\author{Sylvia Ekstr\"om, Georges Meynet and Andr\'e Maeder}
\affil{Geneva Observatory, Maillettes 51, CH - 1290 Sauverny, Switzerland} 

\begin{abstract} 
The effects of rotation on low-metallicity stellar models are twofold: first, the models reach break-up during main sequence and may lose mass by mechanical process; second, strong internal mixing brings freshly synthesized elements towards the surface and raises the effective metallicity to higher values, so that the initially very low radiative winds are enhanced. Those two effects are also found in $Z=0$ models, but to a lesser degree because of structural differences. This weak mass loss becomes interesting though in the case of the black hole-doomed stars (25-140 M$_{\sun}$\ and $>$ 260 M$_{\sun}$) because it allows these stars to contribute to the enrichment of the medium by stellar winds.
\end{abstract}

\section{\boldmath $Z=0$ models}
The effects of rotation on stellar evolution at low $Z$ are presented in detail by Georges Meynet (see contribution in this volume). We come back here on two aspects that play an important role in $Z=0$ stellar models. The results presented here have been obtained by computing a small grid of models with initial abundances $X=0.76$, $Y=0.24$ and $Z=0$, and masses ranging from 15 to 200 M$_{\sun}$. For each mass, we have calculated a rotating and a non-rotating model. The chosen equatorial velocity for the rotating models is $\upsilon_{\mathrm{ini}}=800$ km/s, which may seem fast but corresponds to a similar fraction of the critical velocity than a velocity of 300 km/s in a solar metallicity model. All the models have been computed with the Geneva code, using the same physical ingredients than in \citet{MM8}. For the models reaching $\log L/\mathrm L_{\sun} >$ 6, we have used the radiative mass loss prescription of \citet{kudr02}, with the same adaptations for the $Z=0$ case than in \citet*{marigo03}.

\section{First rotation effect: break-up during main sequence}
\begin{figure}[!ht]
\begin{center}
\includegraphics[height=5cm]{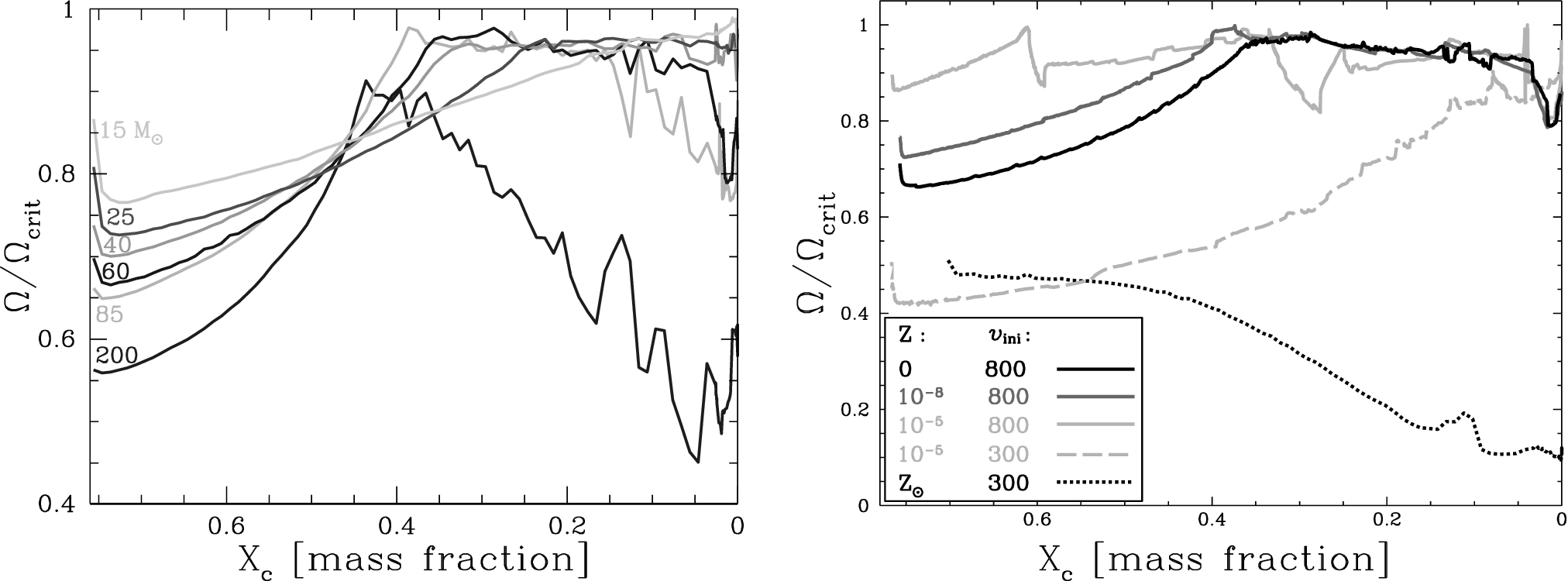}
\caption{Evolution of the $\Omega / \Omega_c$ ratio during the main sequence phase.\protect\\ \textbf{Left}: all the $Z=0$ rotating models. \textbf{Right}: comparison of several 60 M$_{\sun}$ with different metallicities (light grey: $Z=10^{-5}$; middle grey: $Z=10^{-8}$; black: $Z=0$) and different equatorial velocities (continuous: 800 km/s; dashed: 300 km/s). A $Z=\mathrm Z_{\sun}$ stellar model has also been plotted (black dotted) in order to show the effect of strong mass loss on the surface angular velocity.\label{omcrit}}
\end{center}
\end{figure}

Since the models have very low or no radiative winds at all, they have no way to lose angular momentum. During the main sequence (MS) phase, the contracting core accelerates the envelope through meridional circulation coupling and brings the surface to the break-up limit (see Fig. \ref{omcrit}, left).
 As can be seen in this figure, all the $Z=0$ models reach break-up limit. They all remain at break-up until the end of the MS (the 200 and 85 M$_{\sun}$ models seem to depart from break-up, but remain in fact critical if one takes into account the $\Omega \Gamma$ limit). This critical rotation induces a purely mechanical mass loss due to centrifugal force, with a rather weak amplitude because the expelled layers are very superficial and little dense. If we add just a small fraction of metals in the initial abundance, the effect is slightly stronger because the core-envelope coupling is a little more efficient and the models reach break-up earlier in their evolution, as can be seen on Fig. \ref{omcrit} (right). Typically, the 60 M$_{\sun}$ at $Z=0$ loses 3.7\% of its initial mass through this mechanism, while the $Z=10^{-8}$ corresponding model loses 4\% and the $Z=10^{-5}$ model up to 10\%.
 
\section{Second rotation effect: strong mixing during core helium burning}
We have mentioned that the efficiency of the outer cell of the meridional circulation decreases with decreasing $Z$, which leads to steep $\Omega$-gradients and thus a strong internal mixing by shear turbulence. Freshly synthetized elements are thus brought to the surface and the effective metallicity increases.
\begin{figure}[!ht]
\begin{center}
\includegraphics[height=4.7cm]{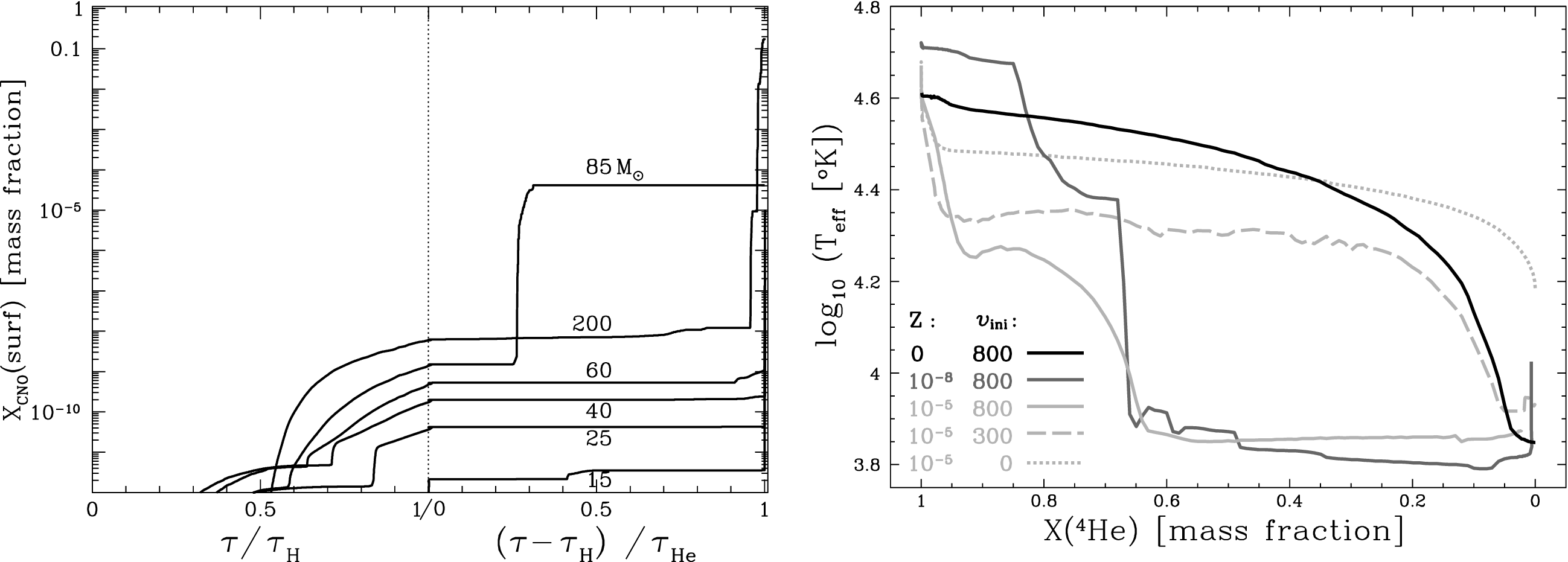}
\caption{\textbf{Left}: evolution of the surface CNO mass fraction with the age of the $Z=0$ rotating models. The age is expressed as a fraction of the MS time ($\tau_{\mathrm H}$) followed by the fraction of $\tau_{\mathrm{He}}$. \textbf{Right}: redward evolution during core helium burning, comparison of several 60 M$_{\sun}$ models with different metallicities (light grey: $Z=10^{-5}$; grey: $Z=10^{-8}$; black: $Z=0$) and different equatorial velocities (cont.: 800 km/s; dashed: 300 km/s; dotted: 0 km/s).\label{surfevol}}
\end{center}
\end{figure}
 As can be seen in Fig. \ref{surfevol} (left), the surface enrichment of the $Z=0$ models begins already during the MS phase and reaches non negligible values only at the very end of the core Helium burning phase. Although the surface metallicity increases then up to $Z_{\mathrm{eff}}=10^{-8}$ or $10^{-4}$ (depending on the model), the amount of mass lost through this mechanism remains modest because the period of time concerned is short. If we again add a small fraction of metals, the effect becomes very important. In fact rotation induces an early redward evolution in these low $Z$ models, as can be seen in Fig. \ref{surfevol} (right). The increase of the effective surface metallicity takes place earlier and the mass removed by this mechanism is very important. For instance, whereas the 60 M$_{\sun}$ at $Z=0$ loses only 0.05\% of its initial mass during this stage, the $Z=10^{-5}$ model loses 28\% and the $Z=10^{-8}$ model up to 56\%. Let us mention that the fact that the $Z=10^{-8}$ model loses more mass than the $Z=10^{-5}$ model is due in part to a blue loop in the HR diagram, at the very end of the core helium burning phase, so the model is brought again at break-up limit \citep[see][]{hl98}.

\section{Total mass loss at \boldmath $Z=0$}
\begin{figure}[!ht]
\begin{center}
\includegraphics[height=5cm]{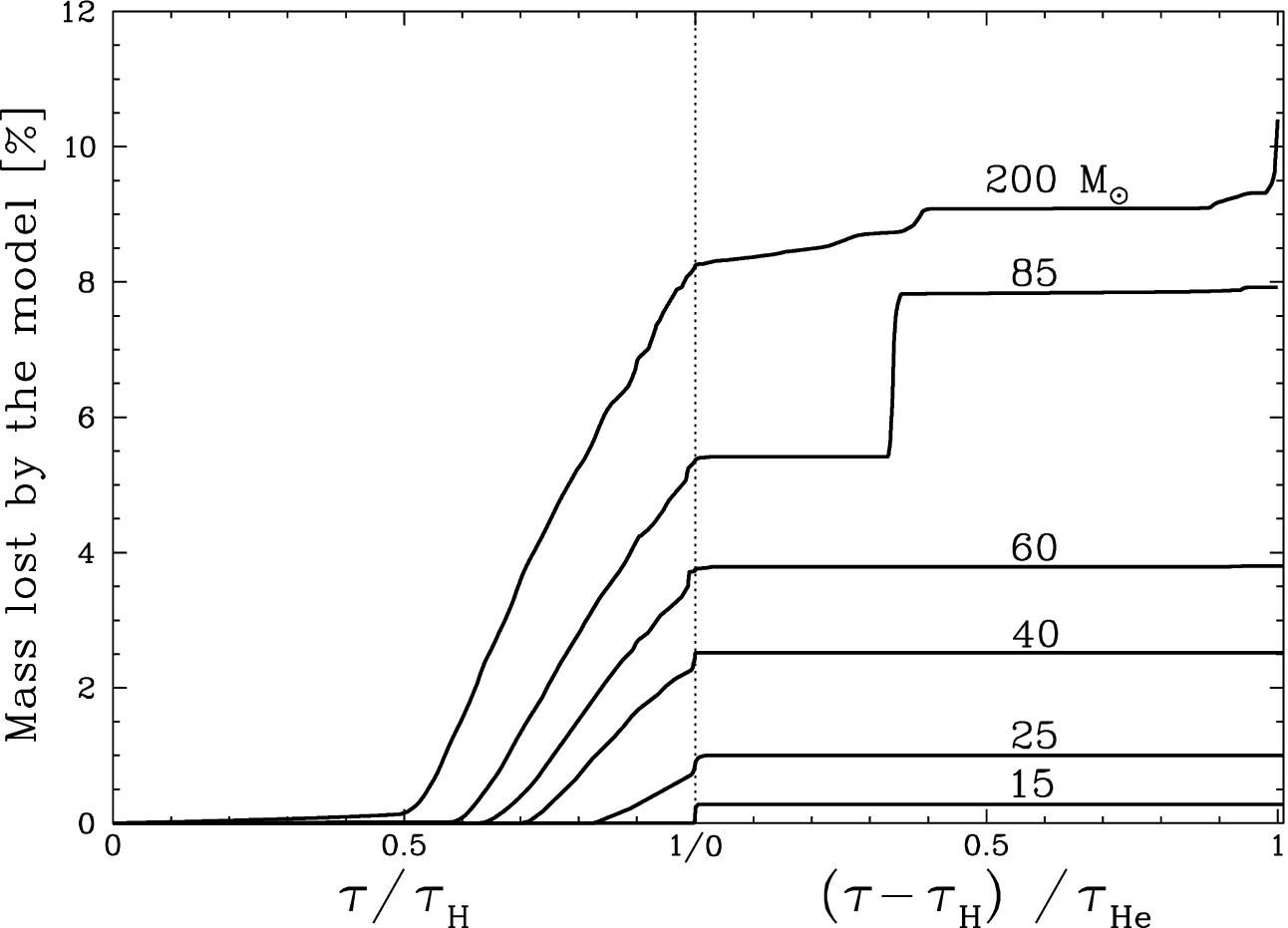}
\caption{Evolution of the mass lost by the $Z=0$ rotating models expressed as the percentage of the initial mass.  The x axis is the same as in Fig. \ref{surfevol} (left). \label{mdot0}}
\end{center}
\end{figure}

If we add both effects, we can see that the total mass lost by the $Z=0$ models remains very modest in the range of masses and velocities considered (see Fig. \ref{mdot0}). The 15 M$_{\sun}$ model loses less than 0.3\% of its initial mass, and the 200 M$_{\sun}$ model only a little more than 10\%. This latter result meets the conclusion of \citet{marigo03}, even though the mass loss found here is a little higher. One can note that for the models ranging up to 40 M$_{\sun}$, the mass loss is almost only due to the MS mechanical process.
 Now, if the star explodes at the end of its life (as a Type II supernova for the low mass end of our sample or as pair-instability supernova for the high mass end) the wind contribution will be totally insignificant in front of the supernova contribution. There is a mass domain where the wind contribution can become non negligible: the mass domain where the stars are supposed to end their life as direct black hole. According to \citet*{h2fwl03}, this mass domain includes stars in the range of 25 to 140 M$_{\sun}$ or above 260 M$_{\sun}$ on the MS. Those stars have long been considered as useless for the chemical evolution of their surroundings. Here we see that they may have a modest, but yet non zero contribution through their winds. If we ponderate the mass ejected by the models with the IMF, and taking into account only the massive stars that may contribute to the chemical evolution at that early time, we see that the part of the wind-only contributors is about 4-9 \% of the total ejected mass, depending on the IMF considered (standard Salpeter or double-peaked as in \citeauthor{nm} \citeyear{nm}). We let the reader judge if this is significant or not, but anyway it's a lower limit because we have not study the case of $M > 260\ \mathrm M_{\sun}$, a range of mass that should experience stronger mass loss.

\section{Discussion}
We have seen that fast rotation enhances mass loss through two mechanisms: the reaching of break-up limit during the main sequence phase and the rising up of the effective metallicity during and after the core helium burning phase. In the strictly $Z=0$ case, both mechanisms play a role but the effects are not very strong and the mass loss remains low for the initial velocity and mass range considered here. Let us mention though that whereas the evolution is not far from constant mass evolution, the rotational mixing experienced by fast rotating models have strong effects on the abundances profiles and consequently on the chemical yields obtained. In particular, the present population III stellar models produce large amounts of primary nitrogen. In the non-zero but very low-metallicity case, we have seen that both mechanisms lead to very strong effects, changing drastically the vision we may have on the extremely metal-poor stars. The results presented here are qualitative but they show a robust trend through a wide range of masses and metallicities and are based on physical ingredients that have more than once given proof of their validity. For a detailed discussion about the subject, we let the reader refer to \citet*{MEM}.

%\acknowledgements %%% Text of acknowledgements runs on after this command.

\end{document}